\documentclass[11pt]{article}
\usepackage{amsmath}
\usepackage{amssymb}
\usepackage{amsfonts}
\usepackage{amscd}
\usepackage{accents}
\usepackage{pst-all}
\usepackage{epsfig}
\usepackage{graphicx}
\usepackage[tight]{subfigure}
\date{\today}
\newcommand{\bmat}{\left(\begin{array}}
\newcommand{\emat}{\end{array}\right)}
\newcommand{\be}{\begin{equation}}
\newcommand{\ee}{\end{equation}}
\newcommand{\ba}{\begin{eqnarray}}
\newcommand{\ea}{\end{eqnarray}}

\def\lsim{\raise0.3ex\hbox{$\;<$\kern-0.75em\raise-1.1ex\hbox{$\sim\;$}}}
\def\gsim{\raise0.3ex\hbox{$\;>$\kern-0.75em\raise-1.1ex\hbox{$\sim\;$}}}

\def\be{\beta}

\begin{document}

\vspace*{-.6in} \thispagestyle{empty}
\begin{flushright}
DESY 12-179
\end{flushright}
\baselineskip = 20pt

\vspace{.5in} {\Large
\begin{center}
{\bf  Metastable Electroweak Vacuum: Implications for Inflation  }
\end{center}}

\vspace{.5in}

\begin{center}
{\bf  Oleg Lebedev and  Alexander Westphal   }  \\

\vspace{.5in}

 \emph{DESY Theory Group, 
Notkestrasse 85, D-22607 Hamburg, Germany
 }
\end{center}

\vspace{.5in}

\begin{abstract}
\noindent
Within the Standard Model, 
the current Higgs and top quark data favor metastability of the electroweak vacuum, although the uncertainties are still significant.  
The true vacuum is many orders of magnitude deeper than ours and the barrier 
separating the two is tiny compared to the depth of the well. This raises 
a cosmological question: how did the Higgs field get trapped in the shallow minimum 
and why did it stay there during inflation? The Higgs initial conditions before 
inflation must be fine--tuned to about one part in $10^8$ in order for the Higgs
field  to end up in the right vacuum.
In this note, we show that these problems can be resolved if there is a small
positive coupling between the Higgs and the inflaton.  
\end{abstract}

The ATLAS and CMS experiments at the LHC have recently observed 
a particle whose properties are well  consistent with those expected 
of the Standard Model Higgs boson. Its mass is determined to be
\begin{eqnarray}
&& M_h = 126.0 \pm 0.4 \pm 0.4 ~ {\rm GeV}~~~, ~~~{\rm ATLAS} \; \cite{:2012gk} \nonumber \\ 
&& M_h = 125.3 \pm 0.4 \pm 0.5 ~ {\rm GeV}~~~, ~~~{\rm CMS} \nonumber \; \cite{:2012gu} \; 
\end{eqnarray}
Such a light Higgs boson coupled with the recent Tevatron top quark mass determination
\begin{eqnarray}
&& m_t^{\rm exp} =173.2 \pm 0.9~{\rm GeV} \;\cite{Lancaster:2011wr} \nonumber
\end{eqnarray}
favors metastability of the electroweak (EW)  vacuum.
 Taking, for example, 
$M_h=125$ GeV and $m_t=173$ GeV, one finds that
the quartic Higgs coupling turns
negative at $\Lambda \sim 10^{10}$ GeV \cite{Degrassi:2012ry}, indicating that 
the electroweak vacuum is not the ground state and therefore only metastable, although 
its lifetime  is greater than the age of the universe \cite{Arnold:1989cb}. 
In fact, the two loop analysis of 
\cite{Degrassi:2012ry} finds that absolute stability is disfavored at 98\% CL for 
$M_h < 126$ GeV, with the main uncertainty coming from the top mass
determination. Although no conclusive statement can yet be made 
\cite{Bezrukov:2012sa,Alekhin:2012py,Holthausen:2011aa}
   as the uncertainties may  be larger than those assumed in \cite{Degrassi:2012ry},
this shows that the current data favor metastability of our vacuum.

\begin{figure}
  \centering
  \includegraphics[height=18em]{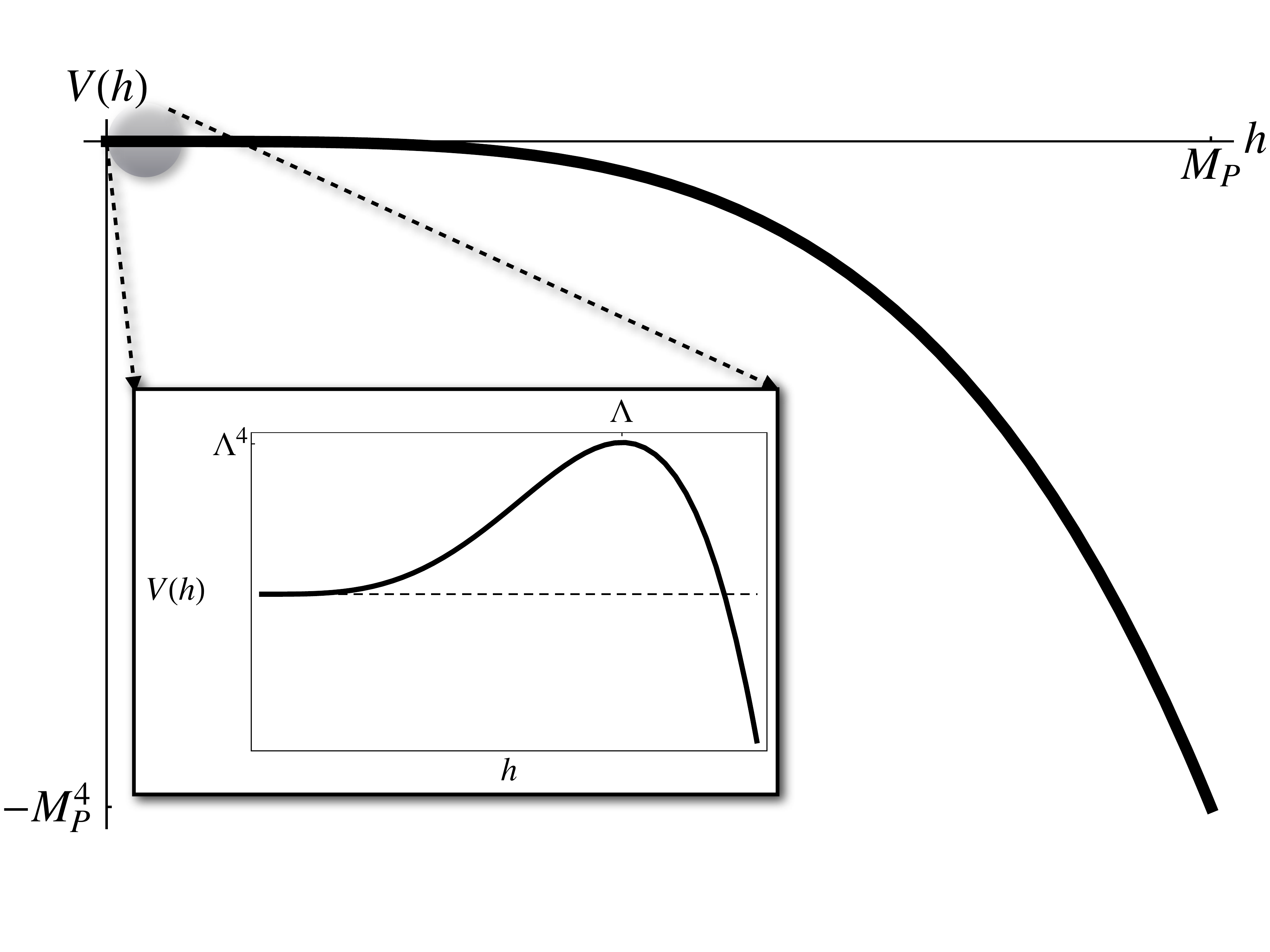}
  \caption{ A $schematic$ view of the Higgs potential ($\Lambda 
\sim 10^{10}$ GeV $\ll { M_{\rm Pl}}$). }
  \label{1}
\end{figure}

Extrapolating the Standard Model all the way to the Planck scale,
one would then conclude  that the Higgs field is trapped in the false vacuum with a much larger 
energy density than that of the ground state and that the barrier separating the two is very 
small compared to the difference of the energy densities (Fig.~\ref{1}):
\begin{equation}
\Lambda^4 \ll M_{\rm Pl}^4 \;. 
\end{equation}
Here we  consider the large Higgs field regime $h \gg v$ ($v=246$ GeV) such that \cite{Altarelli:1994rb}
\begin{equation}
V_{\rm Higgs}(h) \simeq {1\over 4} \lambda (h) ~h^4 \;
\end{equation}
in the unitary gauge,
where $\lambda_h(M_h)\simeq 0.13 $. The coupling
 runs logarithmically with $h$ and turns negative at the ``instability scale'' 
$\Lambda$. 
This raises the cosmological question: how did the universe 
end up in such an energetically disfavored 
state? For generic initial conditions $h \lsim M_{\rm Pl}$ at the beginning of 
inflation \cite{Guth:1980zm,Linde:1983gd}, 
the universe is overwhelmingly likely   
to evolve to the true ground state of the system. 
Not only would that lead to different physics, but would also be catastrophic 
since   the latter is expected to have   a negative Planck--scale energy density and even the 
(COBE--normalized) inflaton
contribution $  10^{-9} M_{\rm Pl}^4$   would not stop the gravitational collapse \cite{Coleman:1980aw},
before thermal effects get a chance to play any role.  
One therefore faces a fine--tuning problem: the initial value of the Higgs field must be 
extremely  small, $h \lsim \Lambda$,  in  Planck units.
For $\Lambda \sim 10^{10}$ GeV, this constitutes a 1 in $10^8$ tuning.\footnote{The Standard Model 
also suffers from another fine--tuning, namely, the hierarchy problem. It is however 
not directly related to the cosmological problem above since the latter has to do with a fine--tuning
in the initial conditions. } 

Furthermore, even if the Higgs field starts at the origin, it will not necessarily remain there
during inflation. Since it is effectively massless, its quantum fluctuations are
of order the Hubble scale $H$. Therefore, for $H > \Lambda$ it is likely to end up in the wrong
vacuum. For $H \ll \Lambda$, it will remain at the origin, yet this does not solve the 
problem of initial conditions.

One possibility to address this issue is based on the landscape idea, along the lines of 
Ref.~\cite{Espinosa:2007qp}. Namely, the regions of multiverse with the ``wrong'' initial
conditions collapse due to the AdS instability such that all the remaining regions would have 
the Higgs field around the origin in field space. However, as shown in \cite{Espinosa:2007qp},
those regions that survive (large--field) inflation would only allow for small curvature perturbations
and the probability of generating the right amount of perturbations is exponentially small.

One may also declare that the Higgs field was prepared in a special state by 
unknown pre--inflationary dynamics, but this simply begs the question.
Possible thermal effects would not do the job since at large $h$ the fields
which couple to the Higgs are heavy and not expected to be in thermal equilibrium.

It is worth noticing that the problem disappears altogether if one allows for 
physics beyond the Standard Model. For example, a tiny coupling of the Higgs
to the hidden sector can stabilize the potential \cite{Lebedev:2012zw} and allow
for ``Higgs--portal'' inflation \cite{Lebedev:2011aq}.
We will however take a conservative view and assume that the SM, with the addition of an inflaton, 
is valid up to the Planck scale. The Higgs itself cannot play the role of an 
inflaton  \cite{Bezrukov:2007ep,Isidori:2007vm,Masina:2011aa,Kamada:2012se}  if the electroweak vacuum is metastable
and the  extra degree of freedom is necessary.

In this work, we show  that the above problems can be resolved if there is a  
Higgs--inflaton coupling which drives the Higgs field to small values during inflation.
Suppose the full scalar potential is given by
\begin{equation} 
V = V_{\rm Higgs}(h) + V_{\rm cross} (h, \phi) + V_{\rm infl}(\phi) \;,
\end{equation}
 where $\phi$ is the inflaton. Then, the Higgs field evolves to the electroweak vacuum
after inflation if 
\begin{equation} 
h_{\rm end} \lsim \Lambda \;,
\end{equation}
where $ h_{\rm end} $ is the Higgs field value  at the end of 
inflation. 
This requirement constraints inflationary models and allowed Higgs--inflaton couplings. 
Restricting ourselves to sub--Planckian  Higgs fields and  using gauge invariance, we can expand 
\begin{equation}
 V_{\rm cross} (h, \phi) = h^2 f_1(\phi) + h^4 f_2 (\phi) + ...
\end{equation}
The desired effect of the cross term is to make the Higgs potential convex and 
induce a Higgs mass term  above the Hubble scale $H$ so that $h$ would evolve
to small values during inflation.

Consider the simplest case of a renormalizable Higgs--inflaton coupling
(as in Higgs--portal models \cite{Silveira:1985rk}) and the quadratic inflaton 
potential in chaotic inflation \cite{Linde:1983gd},
\begin{equation}
V_{\rm cross}  = {1\over 2}  \xi h^2 \phi^2 ~~~,~~~
 V_{\rm infl} = {1\over 2} m^2 \phi^2 \;,
\end{equation}
where $\xi$ is positive. 
The first constraint is that  the coupling $\xi$ should not lead to large radiative corrections
to the inflaton potential during the last 60 $e$--folds. The most important correction is of order (see e.g. \cite{Linde:1990ta})
\begin{equation}  
\Delta V_{\rm infl} \simeq {\xi^2 \over 64 \pi^2 } \; \phi^4 \; \ln {\xi \phi^2 \over m^2}  \;  ,
\end{equation}
so for $m=  10^{-5}$ and $\phi \sim 10$ in $Planck$ units 
 \cite{Lyth:1998xn}, 
the constraint is 
\begin{equation}
\xi \lsim 10^{-6} \;.
\end{equation}
Next, the Higgs potential becomes  dominated by the cross coupling at
\begin{equation}
\phi_0 > \sqrt{ \vert \lambda \vert \over 2 \xi  } ~h_0 \sim 20 \;,
\label{convex}
\end{equation}
where we have taken $\vert \lambda \vert \simeq 10^{-1}$ and chosen the initial Higgs field 
value $h_0 =0.1$ such that higher dimensional Higgs operators are unimportant. 
With these initial conditions, the effective Higgs  mass squared is large and positive, and 
 the field will naturally evolve to small values. Let us consider this process 
in more detail.

\begin{figure}
  \centering
  \includegraphics[height=23em]{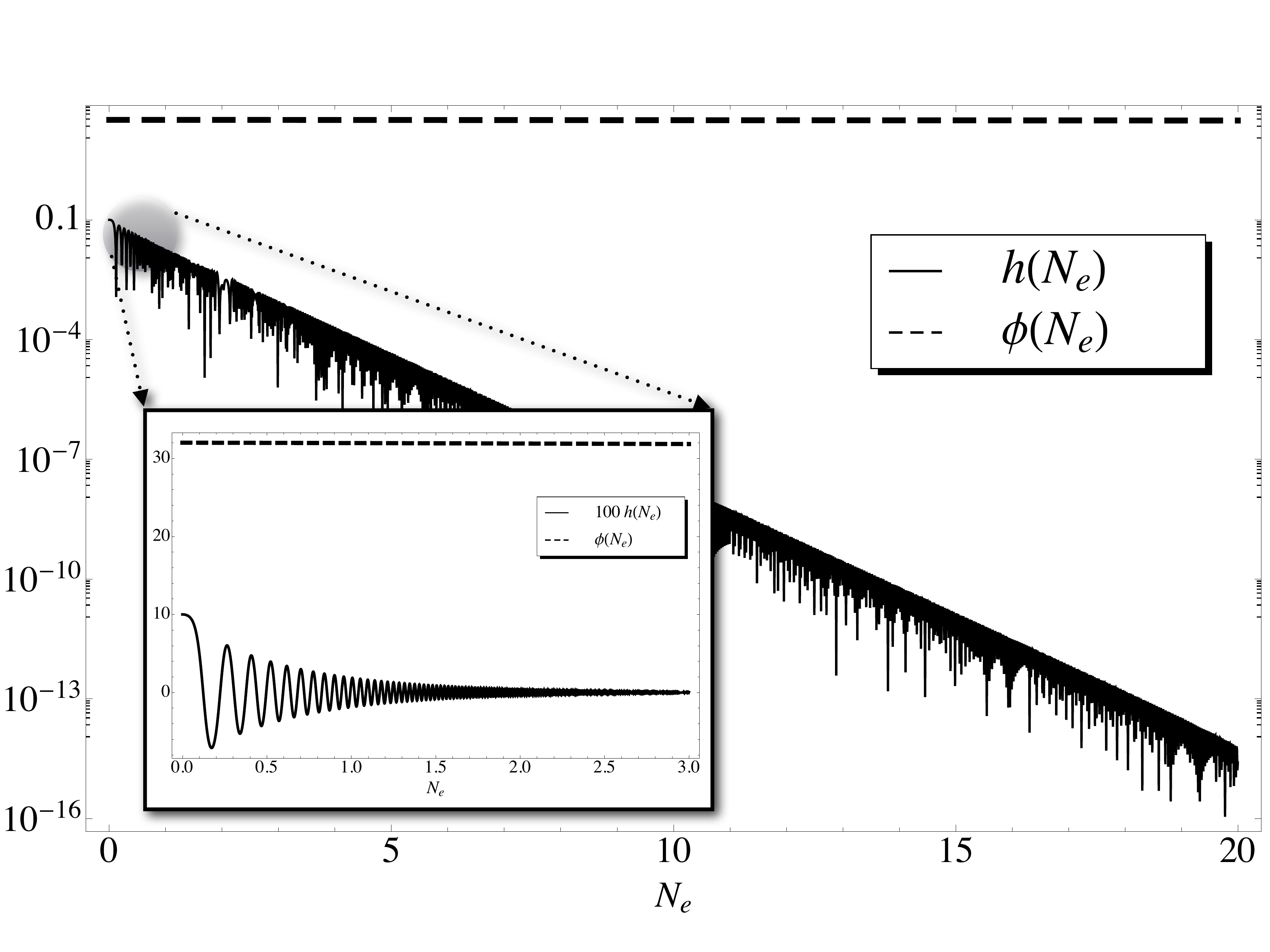}
  \caption{ Evolution of the Higgs field (solid)  and the inflaton (dashed)
as a function of the number of $e$--folds $N_e$. The log--scale plot
shows  the absolute value of $h$ which goes through zero during each
oscillation, but gets cut off at a finite value for numerical reasons. 
The initial values are $\phi_0=32$, $h_0=0.1$ and $\xi=10^{-6}$. }
  \label{2}
\end{figure}

The evolution of the Higgs and inflaton fields is governed by
\begin{eqnarray}
\ddot{h} + 3 H \dot{h} + {\partial V \over \partial h} =0 \;, \nonumber \\
\ddot{\phi} + 3 H \dot{\phi} + {\partial V \over \partial \phi} =0 \;,
\end{eqnarray}
with 
\begin{equation}
3H^2= {1\over 2} \dot{h}^2 + {1\over 2} \dot{\phi}^2 + V 
\end{equation}
and 
\begin{equation}
V \simeq {1\over 2}  \xi h^2 \phi^2 +  {1\over 2} m^2 \phi^2 \;.
\end{equation}
Suppose that initially $\dot{h}$ and $\dot{\phi}$ are insignificant. Then the Hubble
rate is dominated by the cross term, $H_0 \simeq \sqrt{\xi/6} \; \phi_0 h_0$, and
the effective Higgs and inflaton masses satisfy
\begin{equation}
m_\phi \ll H_0 \ll m_h \;.
\label{hierarchy}
\end{equation}
It is then clear that $h$ will evolve quickly leading to a rapid decrease
in the expansion rate,  while the evolution of $\phi$ is
``slow--roll''. At the initial stage of inflation, $h$ evolves according to
\begin{equation}
\ddot{h} + \sqrt{3\over 2}\; \sqrt{\dot{h}^2 + m_h^2 h^2 } \; \dot{h} + m_h^2 h =0 \;,
\end{equation}
with $m_h^2 = \xi \phi_0^2$. The solution to this equation is known in the limit 
$m_h t \gg 1$ (see e.g. late time inflaton evolution  \cite{Mukhanov:2005sc}),
\begin{equation}
h \simeq C \; { \cos m_ht \over m_ht} \;,
\end{equation}
with order one $C$. 
Since $m_h  \simeq 25 H_0 $, the asymptotics $m_h t \gg 1$ is reached
after a few Hubble times $H_0^{-1}$. Therefore, in about  10 Hubble 
times, the amplitude of the Higgs field decreases by 
more than an order of magnitude.
From that point on, the quadratic potential $m^2 \phi^2$ takes over the 
energy density and the usual slow roll inflation takes place.
The expansion rate becomes approximately constant and the Higgs evolution
is governed by 
\begin{equation}
\ddot{h} + 3H \dot{h} + m_h^2 h=0 \;,
\end{equation}
with $H \simeq m \phi_0 / \sqrt{6}$. Its solutions are
$C_{\pm} \exp \big(-3/2 \; H \pm \sqrt{9/4 \; H^2 - m_h^2 }~\big)$.
Since $m_h \gg H$, the  Higgs field decays exponentially,
\begin{equation} 
\vert h \vert \sim  e^{- {3\over 2} H t} \vert h(0) \vert \;.
\end{equation}
Within about 20 $e$-folds, it will be of electroweak size (Fig.~\ref{2}).
On the other hand, the evolution of $\phi$ is ``slow--roll'' and not affected by $h$,
as the latter makes a negligible contribution to the energy density.
The total number of $e$-folds is given approximately by 
$ 1/4 \phi_0^2 > 100$ for $\phi_0$ satisfying (\ref{convex}). 
Finally,
note that the Higgs ``instability'' scale during inflation becomes 
 \begin{equation}
\Lambda \simeq \sqrt{2 \xi \over  \vert \lambda \vert} \; \phi \gg H 
\end{equation} 
and the quantum fluctuations of the Higgs field are irrelevant.

We see that even in the simplest case of $\phi^2$ inflation,
the Higgs--inflaton coupling can stabilize the Higgs potential without 
spoiling the predictions for curvature perturbations. 
During inflation, the Higgs field evolves quickly to small values, yet
the shape of the Higgs potential after inflation is unaffected since  $\xi \ll 1$.
The mechanism is operative in the following range:
\begin{equation}
 10^{-10} \lsim \xi \lsim 10^{-6} \;.
\end{equation}
The upper bound is dictated by the smallness of radiative corrections 
to the inflaton potential,
while the lower bound comes from requiring fast Higgs evolution,
$m_h \gsim H$. The latter is comparable  to the limit on $\xi$
imposed by the dominance of the classical roll of the inflaton over 
quantum fluctuations, 
$\phi_0 \lsim  5/\sqrt{m}$ \cite{Linde:1990ta}.
Note also that for $\xi < 10^{-8}$, the scalar potential is dominated by
$m^2 \phi^2$ and inflation is always ``slow--roll''.

The inflaton-Higgs coupling also provides the reheating mechanism through
 parametric resonance \cite{Kofman:1994rk}. However, whether the reheating 
process is complete or not
depends on the presence of other couplings which make the inflaton unstable.
For instance, $\phi$ can couple to the right--handed 
neutrinos\footnote{The right--handed  neutrinos would be an important ingredient
in the complete framework as they may constitute dark matter \cite{Asaka:2005an}
and/or generate matter--antimatter asymmetry \cite{Fukugita:1986hr}.
Note that  their coupling to the Higgs does not improve the stability of the Higgs potential.}  as
$\phi \bar N N$, or have a trilinear coupling to the Higgs, $\phi h^2$.
The reheating temperature is sensitive to such couplings and no model--independent
prediction can be made. Unless it is  exceedingly high 
($10^{15}$ GeV), the Higgs field will remain at small values throughout the 
reheating  \cite{Espinosa:2007qp}.

As seen from (\ref{convex}), there is no fundamental obstacle to
increase the initial  value of the Higgs field to Planckian values. In that case,
however, calculability is lost due to higher order Higgs operators. 
It is also clear that the mechanism generalizes to other large--field inflationary 
potentials, as long as (\ref{convex}) is satisfied in the slow--roll region
and $m_h \gsim H$.

The inflaton interactions may enjoy the shift symmetry which
can justify the smallness of $\xi$ and higher order operators \cite{Linde:1990ta}.  In particular,  small  values of $\xi$ are radiatively stable and 
not fine--tuned in the t'Hooft
sense since setting $\xi=0$ (and $m_\phi=0$) makes the theory invariant
under $\phi \rightarrow \phi ~+$ const, which is the usual shift symmetry of 
inflationary models. This is in contrast with the fine--tuning in the 
Higgs initial conditions, which is not justified by dynamics and symmetries.

The Higgs coupling to the inflaton obtained above is far too small to be probed at
colliders. This applies to typical inflationary potentials, yet there is a  notable
exception. If one allows for a large non--minimal scalar coupling to gravity as in
\cite{Bezrukov:2007ep}, $\xi$ can be substantial. Suppose we add the term
\begin{equation}
\Delta {\cal L}/\sqrt{-g} ~=~ -{1\over 2} \kappa \phi^2 R \;, 
\end{equation}
where $g$ is the determinant of the metric and $R$ is the scalar curvature. Assume, for
simplicity, that at large $\phi$ the scalar  potential is dominated by
\begin{equation}
V \simeq {1\over 2} \xi h^2 \phi^2 + {1\over 4} \lambda_\phi \phi^4 \;.
\end{equation}
Then, eliminating the non-minimal coupling to gravity by a conformal transformation,
one finds that, for $\kappa \phi^2 \gg 1$,
 the scalar potential in the $\phi$--direction  is exponentially close to a
flat one. Taking $\lambda_\phi \lsim {\cal O}(1)$,  
the correct curvature perturbations are reproduced for $\kappa \sim 10^5$ 
\cite{Bezrukov:2007ep}.
The $h$--direction, on the other hand, is very steep with the effective mass of
order $\sqrt{\xi/\kappa}$, while the Hubble rate is of order  
$\sqrt{\lambda_\phi}/\kappa$ (for details, see \cite{Lebedev:2011aq}). 
Thus, for  $\xi \lsim 0.1$, the Higgs will quickly evolve to small values,
as before. Note that such values of $\xi$ do not lead to significant quantum
corrections  to $\lambda_h$ and the 
inflaton potential. In particular, the electroweak vacuum remains 
metastable. On the other hand, the Higgs coupling to the inflaton is similar in strength
to the Higgs self--coupling, unlike in the previous scenarios. This is the familiar
Higgs portal interaction \cite{Silveira:1985rk}, which, given a light enough inflaton, 
can potentially be probed at colliders. For example, the LHC already places
some constraints on this scenario \cite{Djouadi:2011aa}.

To conclude, we have argued that, in the Standard Model 
(which may include right--handed neutrinos), there is a fine--tuning problem with the 
initial conditions of the Higgs field at the beginning of inflation, if the electroweak
vacuum is indeed metastable. Furthermore, the Higgs field is subject to large
quantum fluctuations during inflation which can destabilize the EW vacuum.
These problems can be circumvented by the presence of 
a small positive coupling of the Higgs to the inflaton, 
$\Delta V = {1\over 2}\xi h^2 \phi^2$, 
as in Higgs portal models. In this case, the Higgs field is driven to small values
during inflation, even if its initial value is close to the Planck scale. 
The important condition is that inflation be ``large--field'', while the specifics of the inflaton
potential are not essential. The coupling $\xi$ can be taken small enough not to affect
the curvature perturbation predictions, in which case it does not change the shape
of the Higgs potential after inflation, either. 
Finally, unlike a fine--tuning of the initial conditions, 
the smallness of the Higgs--inflaton coupling can in principle be justified by an approximate
shift symmetry.

\end{document}